\documentclass[5p]{elsarticle}

\usepackage{multirow}
\usepackage{lineno,hyperref}

\journal{Physics Letters B}

\biboptions{sort&compress}
\newcommand{\fig}[1]{fig.~\ref{#1}}
\newcommand{\tab}[1]{table~\ref{#1}}

\newcommand{\eq}[1]{eq.~(\ref{#1})}

\begin{document}

\begin{frontmatter}

\title{Cross section of $\alpha$-induced reactions on iridium isotopes obtained from thick target yield measurement for the astrophysical $\gamma$ process}

\author[Atomki]{T. Sz\"ucs\corref{mycorrespondingauthor}}
\cortext[mycorrespondingauthor]{Corresponding author}
\ead{szucs.tamas@atomki.mta.hu}

\author[Atomki]{G. G. Kiss}
\author[Atomki]{Gy. Gy\"urky}
\author[Atomki]{Z. Hal\'asz}
\author[Atomki]{Zs. F\"ul\"op}
\author[Basel]{T. Rauscher}

\address[Atomki]{Institute for Nuclear Research (MTA Atomki), 4001 Debrecen, Hungary}
\address[Basel]{Department of Physics, University of Basel, 4052 Basel, Switzerland}

\begin{abstract}
The stellar reaction rates of radiative $\alpha$-capture reactions on heavy isotopes are of crucial importance for the $\gamma$ process network calculations. These rates are usually derived from statistical model calculations, which need to be validated, but the experimental database is very scarce.
This paper presents the results of $\alpha$-induced reaction cross section measurements on iridium isotopes carried out at first close to the astrophysically relevant energy region. Thick target yields of $^{191}$Ir($\alpha$,$\gamma$)$^{195}$Au, $^{191}$Ir($\alpha$,n)$^{194}$Au, $^{193}$Ir($\alpha$,n)$^{196m}$Au, $^{193}$Ir($\alpha$,n)$^{196}$Au reactions have been measured with the activation technique between E$_\alpha = 13.4$\,MeV and 17\,MeV. For the first time the thick target yield was determined with X-ray counting. This led to a previously unprecedented sensitivity.
From the measured thick target yields, reaction cross sections are derived and compared with statistical model calculations. The recently suggested energy-dependent modification of the  $\alpha$+nucleus optical potential gives a good description of the experimental data.
\end{abstract}


\end{frontmatter}

\section{Introduction}
The majority of nuclides heavier than iron are produced via neutron capture reactions \cite{B2FH57, Kappeler11-RMP, 
Arnould07-PR,Thielemann17-ARNPS}. However, there are a few dozens of nuclides on the proton rich side of the valley of stability, the so-called 
$p$-nuclei, which cannot be reached by these processes. These are produced mainly in the so-called $\gamma$ 
process \cite{Rauscher13-RPP}, occurring in hot, dense astrophysical plasmas as encountered, e.g., in core-collapse supernovae 
(ccSN) and thermonuclear supernovae (SNIa). While SNIa remain a promising site for $p$-nucleus production \cite{Travaglio11-AJ,Travaglio15-AJ}, the 
ccSN model calculations still show deficiencies in reproducing the observed $p$-nucleus abundances in some nuclear mass regions 
\cite{Rauscher13-RPP}. Both sites, however, may contribute to the galactic $p$-nucleus content. The deficiencies are partly due to 
the uncertain nuclear physics input \cite{Rauscher16-MNRAS}. The reaction network for the $\gamma$ process involves tens of thousands of 
reactions on thousands of mainly unstable nuclei. The network calculations use mostly theoretical reaction rates calculated with 
the Hauser-Feshbach (H-F) statistical model \cite{Hauser52-PR}. Above neutron number $N=82$ the reaction flow is mainly proceeding 
through chains of ($\gamma$,n) and ($\gamma$,$\alpha$) reactions due to nuclear structure effects (reaction $Q$ values) 
\cite{Rauscher13-RPP,Arnould03-PR}. Experimental reaction rate information can be obtained by measuring the inverse $\alpha$-capture 
reaction cross sections \cite{Mohr07-EPJA,Kiss08-PRL,Rauscher09-PRC} and applying the principle of detailed balance 
\cite{Rauscher11-IJMPE}. Experimental data of $\alpha$-capture reactions in the relevant energy region are still scarce, 
however \cite{Szucs14-NDS}. A comparison of H-F predictions to the scarce low-energy data above $N=82$ have consistently shown an 
overprediction of cross sections \cite{Rauscher13-RPP}. In the astrophysically relevant energy region \cite{Rauscher10-PRC} the H-F cross section 
calculations are only sensitive to the $\alpha$-channel width \cite{Rauscher12-AJS}, which is calculated using global 
$\alpha+$nucleus optical model potentials. Recently, an energy 
dependent modification of the depth of the imaginary part of the widely used McFadden-Satchler potential \cite{McFadden66-NP} was 
shown to describe much better the experimental data \cite{Sauerwein11-PRC, Kiss14-PLB, Kiss15-JPG}. Lately developed further 
alternatives also include energy-dependent modifications of the imaginary part, these are e.g. \cite{Avrigeanu10-PRC, Demetriou02-NPA, 
Mohr13-ADNDT}.

This work presents experimental data for one of the heaviest nuclides investigated so far. For the first time, thick target yield measurement combined with X-ray detection were employed for determining 
$\gamma$-process related cross sections for such a heavy nuclide. The data were compared to H-F calculations 
for further constraining the optical model potential.

\section{Thick target yield and cross section}
Most of the former studies concentrated on the direct measurement of the reaction cross sections. Usually thin 
layers of target material are used, in which the projectile energy loss is small, and by knowing the number of target atoms the 
cross section can be derived at an effective energy. In the present study the projectile stops in the target, therefore, reactions 
take place with all energies between the bombarding energy and zero. Thus the quantity to be measured is the so-called thick target 
yield, i.\,e., the number of reactions per projectile. The number of target atoms is maximized in this way and does not limit the 
yield to be measured. In $\gamma$ process relevant studies the thick target yield technique was applied recently only in the lower 
mass range \cite{Gyurky14-NPA,Fiebiger17-JPG}. This study is the pioneering work in the heavy mass range. 

The thick target yield ($Y_{TT}(E)$) as a function of $\alpha$ energy ($E$) is related to the reaction cross section ($\sigma(E)$) by the following integral formula:
\begin{equation}
Y_{TT}(E) =\int_{0}^{E} \frac{\sigma(E')}{\epsilon_{eff}(E')} dE',
\label{eq:TTY}
\end{equation}
where $\epsilon_{eff}(E)$ is the effective stopping power for the studied isotope, i.e., the stopping power of chemically pure 
iridium divided by the isotopic abundance of the studied isotope. 
From the measured thick target yields the cross section between two energies can be obtained by subtraction:
\begin{equation}
\sigma(E_{eff}) = \frac{\left(Y_{TT}(E_2) - Y_{TT}(E_1)\right)\overline{\epsilon_{eff}}(E_1;E_2)}{E_2 - E_1},
\label{eq:sigma}
\end{equation}
where $\overline{\epsilon_{eff}}(E_1;E_2)$ is the averaged effective stopping power in the $(E_1;E_2)$ energy range.

$E_{eff}$ is determined from the yield curve, per definition 
\begin{equation}
Y_{TT}(E_{eff}) =  \frac{Y_{TT}(E_2) + Y_{TT}(E_1)}{2}.
\label{eq:eff_energy}
\end{equation}

\section{Studied reactions}
Iridium in its natural form consists of two isotopes, $^{191}$Ir and  $^{193}$Ir with 37.3\% and 62.7\% relative abundances, 
respectively. The $\alpha$-induced reactions on both isotopes were investigated in the energy range of E$_\alpha=13.4$\,MeV$-$ $17.0$\,MeV in
0.5\,MeV energy steps. For the investigations the activation technique was used. Therefore only reactions leading to unstable 
nuclei were studied.
The main reaction of interest is the radiative capture of $\alpha$ particles by $^{191}$Ir. Since in the studied energy region the
($\alpha$,n) reaction channel is also open for both isotopes, these reactions were also studied. Although they are not 
immediately important in the $\gamma$ process, their cross sections are mainly sensitive to the $\alpha$-channel width. 
Accordingly they provide an additional constraint for the $\alpha+$nucleus optical potential.

\paragraph{$^{191}$Ir($\alpha$,$\gamma$)$^{195}$Au}
Because there are no previous experimental data for this reaction in the literature, our new data enlarges the existing 
database related to $\gamma$ process reactions.
The reaction product $^{195}$Au has the longest half-life of all studied isotopes (see \tab{tab:param}). For this reaction, not 
only the $\gamma$ rays were used in the $Y_{TT}$ determination, but also the X-rays. However, as the characteristic X-rays 
following the decay of the different studied isotopes are identical, the X-ray peaks in the spectra are initially populated by all 
of them. After several months the reaction products with shorter half-lives decay off and only $^{195}$Au remains. Thus, from 
the time-dependent population of the X-ray peaks, the activity of this isotope can be derived. At the lowest irradiation energies 
the $\gamma$ peak was buried by the background and only the X-rays were strong enough to be visible. The X-ray detection technique 
was tested already for thin targets \cite{Kiss11-PLB, Kiss12-PRC}. In this paper we present the first thick target yield 
measurements via X-ray detection for a $\gamma$ process related study.
\begin{table}[t]
\caption{Decay parameters of the reaction products used for the analysis \cite{NDS194,NDS195,NDS196}.}
\label{tab:param}
\center
\resizebox{0.99\columnwidth}{!}{
\begin{tabular}{l c c r@{.}l @{}l}
Reaction		& Half-life  			& $\gamma$ ray or X-ray& \multicolumn{3}{c}{Intensity} \\
product			& / h				& energy / keV		& \multicolumn{3}{c}{ / \%} \\
\hline
$^{194}$Au 	& 38.02\,{\it10}		& 328.5				&60	& 4	&{\it8}\\
				&					& 293.5				&10	& 58	&{\it15}\\
$^{195}$Au		& 4464.2\,{\it14}	& \phantom{2}66.8	&47	& 2	&{\it11}\\
				&					& \phantom{2}98.9	&11	& 21	&{\it15}\\
$^{196m}$Au	& 9.6\,{\it1}			& 147.8				&43	& 5	&{\it15}\\
				&					& 188.3				&30	& 0	&{\it15}\\
$^{196}$Au		& 148.006\,{\it14}	& 355.7				&87	& 0	&{\it30}\\
				&					& 333.0				&22	& 9	&{\it9}\\
				&					& 426.1				&6	& 6	&{\it3}\\
\end{tabular}
}
\end{table}
\paragraph{$^{191}$Ir($\alpha$,n)$^{194}$Au}
There are two datasets in the literature for this reaction \cite{Bhardwaj92-PRC, Ismail98-Pra}. Both were obtained using the 
stacked foil technique. Only the lowest energy points of these 
studies are within our investigated energy range. Due to the limitations of this technique, however, those points have large 
energy uncertainties.
In our measurement, the energy uncertainty is much smaller even with the subtraction method.

\paragraph{$^{193}$Ir($\alpha$,n)$^{196m}$Au}
The metastable state of $^{196}$Au at an excitation energy of 0.596 MeV has a long enough half-life to be measurable by the 
activation technique. This level decays exclusively by internal transitions to the ground state, producing $\gamma$ rays with high 
relative intensity (see \tab{tab:param}). Using these, the partial thick target yield populating this level was derived. Previously 
in the literature only the ratios of the reaction cross sections leading to the metastable and to the ground state were published, 
and mainly at reaction energies  much higher than our energy range \cite{Gavriluk89-conf, Denisov93-PAN, Chuvilskaya99-BRASP}.

\paragraph{$^{193}$Ir($\alpha$,n)$^{196}$Au}
From the decay of $^{196}$Au the total reaction cross section was derived. Even though $^{196}$Au nuclei in their ground states are also 
produced via the long lived isomeric state, after one day of waiting time the majority of the metastable nuclei 
de-excited. Using only the spectra recorded after this time, the measured decay curve of $^{196}$Au was not distorted. The total 
reaction cross section including the production via the metastable state was calculated from these countings. 

\section{Experimental details}

\paragraph{Targets} For the measurements, 50\,$\mu$m thick high purity (99.9\,\%) iridium foils of natural isotopic composition were used. This 
thickness fulfils the criteria of a thick target to completely stop the $\alpha$ particles. With the maximum energy investigated here (17\,MeV), the average range of an 
$\alpha$ particle in iridium according SRIM \cite{SRIM} is 40\,$\pm$2\,$\mu$m. According to the supplier's specification, the 
iridium foils contain trace amounts of platinum, rhenium, and iron at the ppm level.

\paragraph{Irradiations} For the irradiations, the MGC-20 type cyclotron of Atomki was used. The $\alpha$ particles entered the 
activation chamber through a beam defining aperture and a second aperture was supplied with -300\,V secondary electron suppression 
voltage. The apertures and the chamber were isolated allowing to measure the beam current. The typical $\alpha^{++}$-beam current 
was $2$\,$\mu$A$-2.5$\,$\mu$A. The length of the irradiations was typically $22$\,h$-34$\,h. Since the $\alpha$ particles completely stopped in 
the targets, the possible blistering had to be avoided. Therefore, the irradiation was stopped typically every 12\,h and the 
target was rotated to receive the bombardment at a slightly different spot. With this method, no visible blistering occurred.
The beam current was recorded with a multichannel scaler, stepping the channels every minute. In this way, the small variations 
in beam intensity were followed and taken into account in the data analysis.

\paragraph{$\gamma$-ray and X-ray detection} The produced activity was determined by counting the $\gamma$ and/or X-rays following
the decay of the reaction products (see \tab{tab:param}). For the counting a thin-crystal high-purity germanium detector, a 
so-called Low Energy Photon Spectrometer (LEPS) was used. The detector was equipped with a home-made 4$\pi$ shielding consisting of 
layers of copper, cadmium, and lead \cite{Szucs14-AIPConf}. 

\begin{figure}[t]
\includegraphics[width=0.92\linewidth]{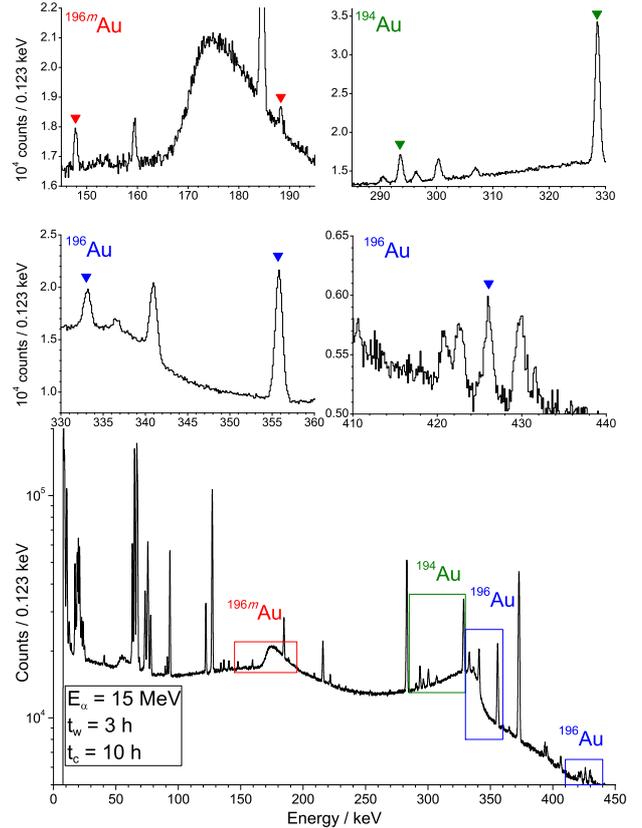}
\caption{\label{fig:gamma} Spectrum of the sample irradiated with 15.0\,MeV $\alpha$ particles. Waiting and counting times are indicated with  $t_w$ and $t_c$, respectively. The upper panels show the peaks used for the activity determination.}
\end{figure}

The activity of the reaction products of the ($\alpha$,n) channels was measured at 3\,cm distance.
The dead time was always below 5\% at the beginning of the counting and decreased to a negligible level within a few hours. 
A typical spectrum taken 3\,h  after the irradiation with E$_\alpha = 15$\,MeV is shown in \fig{fig:gamma}.

\begin{figure}[t]
\includegraphics[width=0.92\columnwidth]{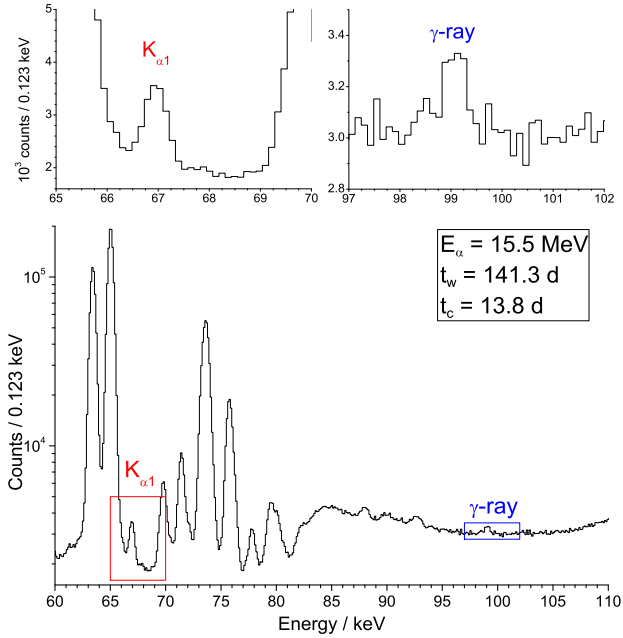}
\caption{\label{fig:Xray} Spectrum of the sample of the 15.5\,MeV irradiation. Waiting and counting times are indicated with  $t_w$ and $t_c$, respectively. The upper panels show the $^{195}$Au K$_{\alpha_1}$ X-ray peak and the $\gamma$ ray peak used for the activity determination.}
\end{figure}

The countings for the $^{191}$Ir($\alpha$,$\gamma$)$^{195}$Au reaction product were done in 1\,cm counting geometry to increase the 
efficiency. Typically these countings were done at least 4 months after the irradiations, hence no notable dead time was experienced 
and only the $^{195}$Au isotope populated the X-ray peaks, i.e., less than 0.5\,\% contribution came from the other isotopes.
In the X-ray spectra the X-ray fluorescence of the bulk iridium was always observed causing peaks at, e.g., 63.3\,keV and 64.9\,keV 
(K$_{\alpha_2}$ and K$_{\alpha_1}$, respectively). 
The X-ray fluorescence was induced by long-lived parasitic activities like $^{57}$Co, which were always procured on the trace 
impurities in the targets.
This kind of fluorescence was not observed in previous thin target measurements \cite{Kiss11-PLB, Kiss12-PRC} because in those cases 
less parasitic activity was produced by the fewer impurity atoms and there was also less material on which the fluorescence could 
be induced.
The K$_{\alpha_2}$ X-ray from the reaction product at 65.1\,keV is buried under the fluorescence peak but thanks to the excellent 
energy resolution of the LEPS detector the K$_{\alpha_1}$ X-ray at 66.8\,keV can be separated from the much more intense 
fluorescence peak. Even when the separation was excellent, the florescence was the main limiting factor of the activity 
determination (see \fig{fig:Xray}). 

\begin{figure}[b]
\includegraphics[width=0.92\columnwidth]{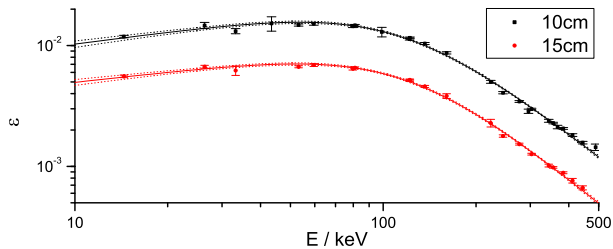}
\caption{\label{fig:eff} The measured detector efficiency at 10\,cm and 15\,cm source to end-cap distance, fitted by an 
exponential function \mbox{$\epsilon(E) = (A E^B + C E^D)^{-1}$} \cite{McFarland91-RR}. One $\sigma$ confidence levels around 
the fit is also shown by dotted lines.}
\end{figure}

The detector efficiency calibration was done with $\gamma$ sources of known activity at end-cap to target distances of 
10\,cm and 15\,cm to avoid true coincidence-summing effects. The obtained efficiency points were fitted with an exponential 
function \cite{McFarland91-RR} as shown in \fig{fig:eff}. At each energy the 1$\sigma$ confidence interval 
of the fit was used for the efficiency uncertainty. 
The efficiency at the actual counting distance (3\,cm) was determined with the help of several targets which were
counted both in 10\,cm and 3\,cm geometry. From the observed count rates, knowing the half-lives of the products and the time 
difference of the countings, the efficiency conversion factors were derived. This factor contains the possible loss due to the true 
coincidence-summing in close geometry. The conversion factors measured with the different sources were consistent. Therefore their 
statistically weighted average was used in the close-geometry efficiency determination. The close-geometry efficiency uncertainty 
contains the uncertainty of the fit and the uncertainty of the conversion factors and thus ranges from 1.5\,\% to 8\,\%. The latter 
value is for the two lines for the metastable state, where the statistical uncertainty in the efficiency ratio measurement 
dominated.

Similar efficiency conversion factors were derived for the 10\,cm to 1\,cm and 15\,cm to 1\,cm counting geometries for the X-ray 
peak and $\gamma$ peak several months after the 17\,MeV irradiations when only the $^{195}$Au reaction product was present in the 
target. Only this source was strong enough for this method, because of the sizeable $^{195}$Au isotope production via the $^{193}$Ir($\alpha$,$2$n)$^{195}$Au reaction. The 1\,cm efficiency was than calculated from both the 10\,cm and 15\,cm 
calibration curves and the weighted average of them was used in the analysis. The final efficiency uncertainty in 1\,cm counting 
distance was 3\,\%.

\section{\label{sec:analysis} Analysis and experimental results}

\paragraph{Thick target yield}
The peaks were fitted by a Gaussian while a linear background was assumed under the peaks. 
The detected counts ($C$) are related to the counting and irradiation parameters as follows,
\begin{equation}
\hspace{-8mm} C = Y_{TT}\,\eta\,I\,\sum_{i=1}^{n}\left(\phi_i\,e^{-(n-i)\,\lambda_x\,\Delta t} \right) e^{-\lambda_x t_w}  \left( 1 - e^{-\lambda_x t_c}  \right)
\end{equation}
where $\eta$ is the absolute detection efficiency, $I$ is the relative intensity of the investigated transition, $\phi_i$ is the incident 
particle flux in the $i$th one minute time window ($\Delta t$) of the multichannel scaler, $\lambda_x$ is the decay constant of the 
given reaction product, $n\,\Delta t$, $t_w$ and $t_c$ are length of the irradiation, the waiting time between the end of the 
irradiation and the beginning of the counting, and the duration of the counting, respectively. 
The spectra were stored on a 1\,h time basis to follow the decay of the reaction products and check the stability of the counting system.
The half-lives of the reaction products were found to be consistent with their literature values. Therefore the spectra were 
summed up to reduce the statistical uncertainty.

Since thick targets were used and activity is created in the bulk of the target, the attenuation of the exiting radiation had to be 
taken into account. To estimate this effect the target was estimated to be built up from 0.01\,$\mu$m thick slices. The attenuation 
of the radiation from each slice was calculated using the known attenuation coefficient of iridium \cite{NIST} and 
averaged weighted by an estimated activity distribution. 
For the calculation of the activity distribution the actual beam energy in each slice was calculated using SRIM \cite{SRIM} and 
considered to be constant within the slice. For each slice cross sections from the NON-SMOKER calculations \cite{Rauscher01-ADNDT} 
were used as the first estimate. Later, the activity distribution was iteratively re-calculated using the obtained final cross 
sections. Note that for the estimation only the energy dependence of the cross section is important and its absolute scale plays no 
role in the activity distribution determination. With higher beam energy the highest attenuation is experienced since the tail of 
the activity distribution penetrates deeper into the sample.
The attenuation of the $\gamma$ rays with energies higher than 200\,keV was less than 0.1\%, for the $\gamma$ rays with energy 188\,keV and 149\,keV was less than 0.5\% and 0.9\%, respectively. The highest attenuation of about 5\% is experienced by the 98.9\,keV $\gamma$ ray. As conservative estimate, 30\% relative uncertainty was assigned to the attenuation.

\begin{table*}[t]
\caption{Experimental thick target yields ($10^{-12}$ reactions / incident particle). $i$: At the marked energies the $^{195}$Au 
activity was also created through the $^{193}$Ir($\alpha$,$2$n)$^{195}$Au reaction channel.}
\label{tab:TTY}
\centering
\begin{tabular}{l c c c c}
E$_\alpha$ / MeV& $^{191}$Ir($\alpha$,n)$^{194}$Au	& $^{191}$Ir($\alpha$,$\gamma$)$^{195}$Au	& $^{193}$Ir($\alpha$,n)$^{196m}$Au	& $^{193}$Ir($\alpha$,n)$^{196}$Au \\
\hline
17.00 $\pm$	0.05	&	13200	$\pm$	500	&	$i$	&	310	$\pm$	20	&	21100	$\pm$	900	\\
16.50 $\pm$	0.05	&	6600	$\pm$	200	&	$i$	&	146	$\pm$	10	&	11000	$\pm$	500	\\
16.00 $\pm$	0.05	&	2240	$\pm$	80	&	10.9	$\pm$	0.6		&	42	$\pm$	2	&	3670	$\pm$	160	\\
15.50 $\pm$	0.05	&	770		$\pm$	30	&	3.4		$\pm$	0.2		&	13.1$\pm$	0.9	&	1310	$\pm$	60	\\
15.00 $\pm$	0.05	&	243		$\pm$	8	&	1.53	$\pm$	0.19	&	3.4	$\pm$	0.3	&	410		$\pm$	18	\\
14.50 $\pm$	0.04	&	77		$\pm$	3	&	0.76	$\pm$	0.05	&	0.96$\pm$	0.13&	136		$\pm$	6	\\
14.00 $\pm$	0.04	&	12.4	$\pm$	0.5	&	0.24	$\pm$	0.04	&		$<$	0.21	&	20.7	$\pm$	1.0	\\
13.40 $\pm$	0.04	&	4.9		$\pm$	0.2	&			$<$		0.47	&		$<$	0.16	&	7.0		$\pm$	0.4	\\
\end{tabular}
\end{table*}

Thick target yields for each ($\alpha$,n) reaction channel were determined from more than one $\gamma$ peak and consistent results
were found. In case of the $^{191}$Ir($\alpha$,$\gamma$)$^{195}$Au reaction the $Y_{TT}$ above 15.5\,MeV were determined both from 
the detected X-rays and $\gamma$ rays. Again, consistent results  within their statistical uncertainty were found. Below 15.5\,MeV the $\gamma$ peak was not visible, 
thus only the X-ray was used for the $Y_{TT}$ determination.

The $Y_{TT}$ obtained from the different transitions were averaged using their statistical weight, which is the combination of the 
uncertainty of the fitted peak area, the uncertainty of the relative intensity of the given peak, the efficiency uncertainty, and 
the uncertainty of the attenuation. After the averaging, the uncertainty of the absolute intensity per decay and the beam current 
uncertainty (3\,\%) were quadratically added. The obtained $Y_{TT}$ are shown in \tab{tab:TTY}.

Above 16.01\,MeV, the $^{193}$Ir($\alpha$,2n)$^{195}$Au reaction channel is open, producing the same isotope as 
$^{191}$Ir($\alpha$,$\gamma$)$^{195}$Au. Therefore, no thick target yields were determined for the radiative capture at 16.5\,MeV 
and 17\,MeV. 

\paragraph{Cross sections}
An average cross section between two energies was derived from the thick target yield using \eq{eq:sigma}. The differentiation of the $Y_{TT}$ has
been done  for each transition using the statistical error only. After that the relative uncertainty of the 
intensity of the given peak, detection efficiency and attenuation was quadratically added to the relative uncertainty of the derived 
cross sections. The consistent cross section values were then averaged using these uncertainties.
Finally, the uncertainty of the absolute intensity per decay, the beam current, 
and the stopping power uncertainty (4\,\%) were quadratically added to the relative uncertainty of the averaged value. 

For the effective energy determination, an exponential curve was fitted to the measured yield points. The quoted effective energy 
was calculated by \eq{eq:eff_energy}. The energy error contains the beam energy uncertainty of 0.3\,\% and an additional 0.5\,\%
uncertainty, which accounts for the considered energy dependence and fit uncertainty of the yield. The derived cross sections are 
shown in \tab{tab:XS} and in the figures later.

\begin{table*}[t]
\caption{Derived reaction cross sections in $\mu$barn.}
\label{tab:XS}
\center
\begin{tabular}{c c}
\multirow{8}{*}{
\begin{tabular}{c c c}
E$_{eff_{c.m.}}$ / MeV	& $^{191}$Ir($\alpha$,n)$^{194}$Au	& $^{191}$Ir($\alpha$,$\gamma$)$^{195}$Au \\ 
\hline
16.47 $\pm$	0.10&	1420 $\pm$	75	&	--	\\
15.98 $\pm$	0.10&	956	$\pm$	50	&	--	\\
15.49 $\pm$	0.09&	329	$\pm$	17	&	1.69 $\pm$	0.14	\\
15.00 $\pm$	0.09&	119	$\pm$	6	&	0.46 $\pm$	0.07	\\
14.51 $\pm$	0.09&	39.0 $\pm$	2.1	&	0.19 $\pm$	0.05	\\
14.02 $\pm$	0.08&	15.3 $\pm$	0.8	&	0.08 $\pm$	0.02	\\
13.50 $\pm$	0.08&	1.50 $\pm$	0.09&		$<$		0.05
\end{tabular}
} &  
\multirow{8}{*}{
\begin{tabular}{c c c}
E$_{eff_{c.m.}}$ / MeV	& $^{193}$Ir($\alpha$,n)$^{196m}$Au	& $^{193}$Ir($\alpha$,n)$^{196}$Au \\ 
\hline
16.48 $\pm$	0.10&	20.4 $\pm$	1.7	&	1295$\pm$	76	\\
15.99 $\pm$	0.10&	13.4$\pm$	1.1	&	952	$\pm$	56	\\
15.50 $\pm$	0.09&	3.9	$\pm$	0.3	&	313 $\pm$	18	\\
15.01 $\pm$	0.09&	1.33$\pm$	0.12&	121 $\pm$	7	\\
14.52 $\pm$	0.09&	0.34$\pm$	0.04&	37.8 $\pm$	2.3	\\
14.03 $\pm$	0.08&	0.12$\pm$	0.03&	16.3 $\pm$	1.0	\\
13.51 $\pm$	0.08&		$<$	0.02	&	1.64 $\pm$	0.10
\end{tabular}
} \\ 
\\ 
\\ 
\\ 
\\ 
\\ 
\\ 
\end{tabular}
\end{table*}

\section{Discussion}
The experimental data have been compared with sta\-tis\-ti\-cal-model calculations performed with the SMARAGD code \cite{smaragd}.

\begin{figure}[t]
\includegraphics[width=0.99\columnwidth]{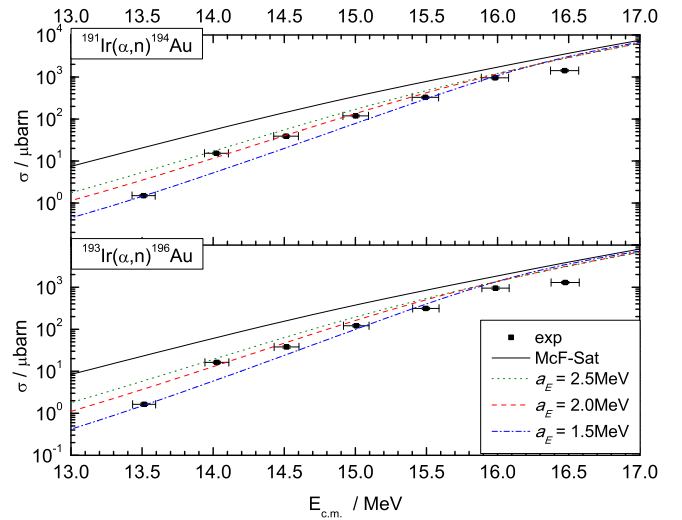}
\caption{\label{fig:an} $^{191}$Ir($\alpha$,n)$^{194}$Au and $^{193}$Ir($\alpha$,n)$^{196}$Au reaction cross sections compared with statistical model calculations. The black dots are the experimental data. Black solid, green dotted, red dashed, and blue dot-dashed lines are the calculations with the standard McFadden-Satchler potential, and with the modified potential with $a_E = 2.5, 2.0, 1.5$\,MeV, respectively.}
\end{figure}

The ($\alpha$,n) cross sections are solely sensitive to the $\alpha$-channel width \cite{Rauscher12-AJS}. 
As can be seen in \fig{fig:an}, the standard McFadden-Satchler potential \cite{McFadden66-NP} does not reproduce well the measured 
data. A good reproduction is found when using the modified energy-dependent potential of \cite{Sauerwein11-PRC}. In this approach 
the energy-dependent depth of the imaginary part is given by
\begin{equation}
W(C,E_\mathrm{c.m.}^\alpha)=\frac{25}{1+e^{\left(0.9E_\mathrm{C}-E_\mathrm{c.m.}^\alpha \right)/a_E}} \quad \mathrm{MeV},
\end{equation}
where $E_\mathrm{C}$ is the height of the Coulomb barrier. Choosing $a_E=2.0$\,MeV gives the best description of the present 
experimental data.

\begin{figure}[htb]
\includegraphics[width=0.99\columnwidth]{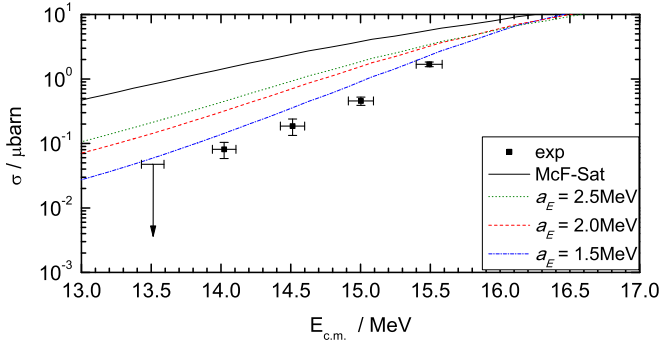}
\caption{\label{fig:191ag} Same as \fig{fig:an} but for the $^{191}$Ir($\alpha$,$\gamma$)$^{195}$Au reaction.}
\end{figure}

Using the same $\alpha$-channel width for calculating the ($\alpha$,$\gamma$) cross section the model overestimates the experimental data (see
\fig{fig:191ag}). In this reaction channel, the calculated cross sections above the ($\alpha$,n) threshold are equally sensitive 
to the $\alpha$-, neutron-, and $\gamma$-widths. Since the $\alpha$ width has been determined by the ($\alpha$,n) 
reaction, the poor reproduction has to be ascribed to the neutron- and/or $\gamma$-widths. These, on the other hand, do not play a 
role to determine the astrophysical reaction rates involving $\alpha$ particles in the $\gamma$ process.

\section{Summary}
Thick target yields of $\alpha$-induced reactions on iridium of natural isotopic composition were measured in the energy range of 
$E_\alpha = 13.4$\,MeV and 17\,MeV with the activation method. The combination of X-ray detection with thick target yield measurements 
has been performed in this mass region for the first time, allowing to measure the reaction cross sections at lower energies than 
ever before. From the measured thick target yields, reaction cross sections were derived and 
compared with statistical model calculations. The results show that the recently suggested energy-dependent modification of the 
widely used McFadden-Satchler $\alpha$+nucleus optical potential gives a good description of the experimental data. The 
$\gamma$- and neutron widths above the ($\alpha$,n) threshold cannot be further constrained by the present data but are not 
relevant for the astrophysical $\gamma$ process.

\section*{Acknowledgement}

This work was supported by NKFIH (K108459, K120666) and the EU COST Action CA16117 (ChETEC).

\bibliography{Ir_alpha_capt}

\begin{thebibliography}{42}
\expandafter\ifx\csname natexlab\endcsname\relax\def\natexlab#1{#1}\fi
\providecommand{\url}[1]{\texttt{#1}}
\providecommand{\href}[2]{#2}
\providecommand{\path}[1]{#1}
\providecommand{\DOIprefix}{doi:}
\providecommand{\ArXivprefix}{arXiv:}
\providecommand{\URLprefix}{URL: }
\providecommand{\Pubmedprefix}{pmid:}
\providecommand{\doi}[1]{\href{http://dx.doi.org/#1}{\path{#1}}}
\providecommand{\Pubmed}[1]{\href{pmid:#1}{\path{#1}}}
\providecommand{\bibinfo}[2]{#2}
\ifx\xfnm\relax \def\xfnm[#1]{\unskip,\space#1}\fi
\bibitem[{Burbidge et~al.(1957)Burbidge, Burbidge, Fowler, and Hoyle}]{B2FH57}
\bibinfo{author}{E.~M. Burbidge}, \bibinfo{author}{G.~R. Burbidge},
  \bibinfo{author}{W.~A. Fowler}, \bibinfo{author}{F.~Hoyle},
  \bibinfo{journal}{Rev. Mod. Phys.} \bibinfo{volume}{29}
  (\bibinfo{year}{1957}) \bibinfo{pages}{547--650}.
\bibitem[{K\"appeler et~al.(2011)K\"appeler, Gallino, Bisterzo, and
  Aoki}]{Kappeler11-RMP}
\bibinfo{author}{F.~K\"appeler}, \bibinfo{author}{R.~Gallino},
  \bibinfo{author}{S.~Bisterzo}, \bibinfo{author}{W.~Aoki},
  \bibinfo{journal}{Rev. Mod. Phys.} \bibinfo{volume}{83}
  (\bibinfo{year}{2011}) \bibinfo{pages}{157--193}.
\bibitem[{Arnould et~al.(2007)Arnould, Goriely, and Takahashi}]{Arnould07-PR}
\bibinfo{author}{M.~Arnould}, \bibinfo{author}{S.~Goriely},
  \bibinfo{author}{K.~Takahashi}, \bibinfo{journal}{Phys. Rep.}
  \bibinfo{volume}{450} (\bibinfo{year}{2007}) \bibinfo{pages}{97--213}.
\bibitem[{Thielemann et~al.(2017)Thielemann, Eichler, Panov, and
  Wehmeyer}]{Thielemann17-ARNPS}
\bibinfo{author}{F.-K. Thielemann}, \bibinfo{author}{M.~Eichler},
  \bibinfo{author}{I.~Panov}, \bibinfo{author}{B.~Wehmeyer},
  \bibinfo{journal}{Annu. Rev. Nucl. Part. Sci.} \bibinfo{volume}{67}
  (\bibinfo{year}{2017}) \bibinfo{pages}{253--274}.
\bibitem[{Rauscher et~al.(2013)Rauscher, Dauphas, Dillmann, Fr\"ohlich,
  F\"ul\"op, and Gy\"urky}]{Rauscher13-RPP}
\bibinfo{author}{T.~Rauscher}, et~al., \bibinfo{journal}{Rep. Prog. Phys.}
  \bibinfo{volume}{76} (\bibinfo{year}{2013}) \bibinfo{pages}{066201}.
\bibitem[{Travaglio et~al.(2011)Travaglio, R\"opke, Gallino, and
  Hillebrandt}]{Travaglio11-AJ}
\bibinfo{author}{C.~Travaglio}, \bibinfo{author}{F.~K. R\"opke},
  \bibinfo{author}{R.~Gallino}, \bibinfo{author}{W.~Hillebrandt},
  \bibinfo{journal}{Astrophys. J.} \bibinfo{volume}{739} (\bibinfo{year}{2011})
  \bibinfo{pages}{93}.
\bibitem[{Travaglio et~al.(2015)Travaglio, Gallino, Rauscher, R\"opke, and
  Hillebrandt}]{Travaglio15-AJ}
\bibinfo{author}{C.~Travaglio}, et~al., \bibinfo{journal}{Astrophys. J.}
  \bibinfo{volume}{799} (\bibinfo{year}{2015}) \bibinfo{pages}{54}.
\bibitem[{Rauscher et~al.(2016)Rauscher, Nishimura, Hirschi, Cescutti, Murphy,
  and Heger}]{Rauscher16-MNRAS}
\bibinfo{author}{T.~Rauscher}, et~al., \bibinfo{journal}{Mon. Not. R. Astron.
  Soc.} \bibinfo{volume}{463} (\bibinfo{year}{2016})
  \bibinfo{pages}{4153}.
\bibitem[{Hauser and Feshbach(1952)}]{Hauser52-PR}
\bibinfo{author}{W.~Hauser}, \bibinfo{author}{H.~Feshbach},
  \bibinfo{journal}{Phys. Rev.} \bibinfo{volume}{87} (\bibinfo{year}{1952})
  \bibinfo{pages}{366--373}.
\bibitem[{Arnould and Goriely(2003)}]{Arnould03-PR}
\bibinfo{author}{M.~Arnould}, \bibinfo{author}{S.~Goriely},
  \bibinfo{journal}{Phys. Rep.} \bibinfo{volume}{384} (\bibinfo{year}{2003})
  \bibinfo{pages}{1--84}.
\bibitem[{Mohr et~al.(2007)Mohr, F\"ul\"op, and Utsunomiya}]{Mohr07-EPJA}
\bibinfo{author}{P.~Mohr}, \bibinfo{author}{Z.~F\"ul\"op},
  \bibinfo{author}{H.~Utsunomiya}, \bibinfo{journal}{Eur. Phys. J. A}
  \bibinfo{volume}{32} (\bibinfo{year}{2007}) \bibinfo{pages}{357--369}.
\bibitem[{Kiss et~al.(2008)Kiss, Rauscher, Gy\"urky, Simon, F\"ul\"op, and
  Somorjai}]{Kiss08-PRL}
\bibinfo{author}{G.~G. Kiss}, et~al., \bibinfo{journal}{Phys. Rev. Lett.}
  \bibinfo{volume}{101} (\bibinfo{year}{2008}) \bibinfo{pages}{191101}.
\bibitem[{Rauscher et~al.(2009)Rauscher, Kiss, Gy\"urky, Simon, F\"ul\"op, and
  Somorjai}]{Rauscher09-PRC}
\bibinfo{author}{T.~Rauscher}, et~al., \bibinfo{journal}{Phys. Rev. C}
  \bibinfo{volume}{80} (\bibinfo{year}{2009}) \bibinfo{pages}{035801}.
\bibitem[{Rauscher(2011)}]{Rauscher11-IJMPE}
\bibinfo{author}{T.~Rauscher}, \bibinfo{journal}{Int. J Mod. Phys. E}
  \bibinfo{volume}{20} (\bibinfo{year}{2011}) \bibinfo{pages}{1071--1169}.
\bibitem[{Sz\"ucs et~al.(2014)Sz\"ucs, Dillmann, Plag, and
  F\"ul\"op}]{Szucs14-NDS}
\bibinfo{author}{T.~Sz\"ucs}, \bibinfo{author}{I.~Dillmann},
  \bibinfo{author}{R.~Plag}, \bibinfo{author}{Z.~F\"ul\"op},
  \bibinfo{journal}{Nucl. Data Sheets} \bibinfo{volume}{120}
  (\bibinfo{year}{2014}) \bibinfo{pages}{191--193}.
  \bibinfo{note}{http://www.kadonis.org/pprocess}.
\bibitem[{Rauscher(2010)}]{Rauscher10-PRC}
\bibinfo{author}{T.~Rauscher}, \bibinfo{journal}{Phys. Rev. C}
  \bibinfo{volume}{81} (\bibinfo{year}{2010}) \bibinfo{pages}{045807}.
\bibitem[{Rauscher(2012)}]{Rauscher12-AJS}
\bibinfo{author}{T.~Rauscher}, \bibinfo{journal}{Astrophys. Jour. Suppl. Ser.}
  \bibinfo{volume}{201} (\bibinfo{year}{2012}) \bibinfo{pages}{26}.
\bibitem[{McFadden and Satchler(1966)}]{McFadden66-NP}
\bibinfo{author}{L.~McFadden}, \bibinfo{author}{G.~Satchler},
  \bibinfo{journal}{Nuclear Physics} \bibinfo{volume}{84}
  (\bibinfo{year}{1966}) \bibinfo{pages}{177--200}.
\bibitem[{Sauerwein et~al.(2011)Sauerwein, Becker, Dombrowski, Elvers, Endres,
  Giesen, Hasper, Hennig, Netterdon, Rauscher, Rogalla, Zell, and
  Zilges}]{Sauerwein11-PRC}
\bibinfo{author}{A.~Sauerwein}, et~al., \bibinfo{journal}{Phys. Rev. C}
  \bibinfo{volume}{84} (\bibinfo{year}{2011}) \bibinfo{pages}{045808}.
\bibitem[{Kiss et~al.(2014)Kiss, Sz\"ucs, Rauscher, T\"or\"ok, F\"ul\"op,
  Gy\"urky, Hal{\'{a}}sz, and Somorjai}]{Kiss14-PLB}
\bibinfo{author}{G.~G. Kiss}, et~al., \bibinfo{journal}{Phys. Lett. B}
  \bibinfo{volume}{735} (\bibinfo{year}{2014}) \bibinfo{pages}{40--44}.
\bibitem[{Kiss et~al.(2015)Kiss, Sz\"ucs, Rauscher, T\"or\"ok, Csedreki,
  F\"ul\"op, Gy\"urky, and Hal{\'{a}}sz}]{Kiss15-JPG}
\bibinfo{author}{G.~G. Kiss}, et~al., \bibinfo{journal}{J. Phys. G: Nucl. Part.
  Phys.} \bibinfo{volume}{42} (\bibinfo{year}{2015}) \bibinfo{pages}{055103}.
\bibitem[{Avrigeanu and Avrigeanu(2010)}]{Avrigeanu10-PRC}
\bibinfo{author}{M.~Avrigeanu}, \bibinfo{author}{V.~Avrigeanu},
  \bibinfo{journal}{Phys. Rev. C} \bibinfo{volume}{82} (\bibinfo{year}{2010})
  \bibinfo{pages}{014606}.
\bibitem[{Demetriou et~al.(2002)Demetriou, Grama, and
  Goriely}]{Demetriou02-NPA}
\bibinfo{author}{P.~Demetriou}, \bibinfo{author}{C.~Grama},
  \bibinfo{author}{S.~Goriely}, \bibinfo{journal}{Nucl. Phys. A}
  \bibinfo{volume}{707} (\bibinfo{year}{2002}) \bibinfo{pages}{253--276}.
\bibitem[{Mohr et~al.(2013)Mohr, Kiss, F\"ul\"op, Galaviz, Gy\"urky, and
  Somorjai}]{Mohr13-ADNDT}
\bibinfo{author}{P.~Mohr}, et~al., \bibinfo{journal}{At. Data Nucl. Data
  Tables} \bibinfo{volume}{99} (\bibinfo{year}{2013})
  \bibinfo{pages}{651--679}.
\bibitem[{Gy\"urky et~al.(2014)Gy\"urky, Vakulenko, F\"ul\"op, Hal{\'{a}}sz,
  Kiss, Somorjai, and Sz\"ucs}]{Gyurky14-NPA}
\bibinfo{author}{Gy.~Gy\"urky}, et~al., \bibinfo{journal}{Nucl. Phys. A}
  \bibinfo{volume}{922} (\bibinfo{year}{2014}) \bibinfo{pages}{112--125}.
\bibitem[{Fiebiger et~al.(2017)Fiebiger, Slavkovsk\'a, Giesen, G\"obel,
  Heftrich, Heiske, Reifarth, Schmidt, Sonnabend, Thomas, and
  Weigand}]{Fiebiger17-JPG}
\bibinfo{author}{S.~Fiebiger}, et~al., \bibinfo{journal}{J. Phys. G: Nucl.
  Part. Phys.} \bibinfo{volume}{44} (\bibinfo{year}{2017})
  \bibinfo{pages}{075101}.
\bibitem[{Kiss et~al.(2011)Kiss, Rauscher, Sz\"ucs, Kert\'esz, F\"ul\"op,
  Gy\"urky, Fr\"ohlich, Farkas, Elekes, and Somorjai}]{Kiss11-PLB}
\bibinfo{author}{G.~G. Kiss}, et~al., \bibinfo{journal}{Phys. Lett. B}
  \bibinfo{volume}{695} (\bibinfo{year}{2011}) \bibinfo{pages}{419--423}.
\bibitem[{Kiss et~al.(2012)Kiss, Sz\"ucs, T\"or\"ok, Korkulu, Gy\"urky,
  Hal\'asz, F\"ul\"op, Somorjai, and Rauscher}]{Kiss12-PRC}
\bibinfo{author}{G.~G. Kiss}, et~al., \bibinfo{journal}{Phys. Rev. C}
  \bibinfo{volume}{86} (\bibinfo{year}{2012}) \bibinfo{pages}{035801}.
\bibitem[{Singh(2006)}]{NDS194}
\bibinfo{author}{B.~Singh}, \bibinfo{journal}{Nuclear Data Sheets}
  \bibinfo{volume}{107} (\bibinfo{year}{2006}) \bibinfo{pages}{1531--1746}.
\bibitem[{Huang and Kang(2014)}]{NDS195}
\bibinfo{author}{X.~Huang}, \bibinfo{author}{M.~Kang},
  \bibinfo{journal}{Nuclear Data Sheets} \bibinfo{volume}{121}
  (\bibinfo{year}{2014}) \bibinfo{pages}{395--560}.
\bibitem[{Xiaolong(2007)}]{NDS196}
\bibinfo{author}{H.~Xiaolong}, \bibinfo{journal}{Nuclear Data Sheets}
  \bibinfo{volume}{108} (\bibinfo{year}{2007}) \bibinfo{pages}{1093--1286}.
\bibitem[{Bhardwaj et~al.(1992)Bhardwaj, Rizvi, and Chaubey}]{Bhardwaj92-PRC}
\bibinfo{author}{M.~K. Bhardwaj}, \bibinfo{author}{I.~A. Rizvi},
  \bibinfo{author}{A.~K. Chaubey}, \bibinfo{journal}{Phys. Rev. C}
  \bibinfo{volume}{45} (\bibinfo{year}{1992}) \bibinfo{pages}{2338--2342}.
\bibitem[{Ismail(1998)}]{Ismail98-Pra}
\bibinfo{author}{M.~Ismail}, \bibinfo{journal}{Pramana} \bibinfo{volume}{50}
  (\bibinfo{year}{1998}) \bibinfo{pages}{173--189}.
\bibitem[{Gavriluk et~al.(1989)Gavriluk, Zheltonojskij, Stepanenko, and
  Kharlamov}]{Gavriluk89-conf}
\bibinfo{author}{V.~Gavriluk}, \bibinfo{author}{V.~Zheltonojskij},
  \bibinfo{author}{V.~Stepanenko}, \bibinfo{author}{V.~Kharlamov},
  \bibinfo{booktitle}{39.Conf. Nucl. Spectrosc. and Nucl. Struct.,Tashkent}, p.
  \bibinfo{pages}{374}.
\bibitem[{Denisov et~al.(1993)Denisov, Zheltonozhskii, and
  Reshit'ko}]{Denisov93-PAN}
\bibinfo{author}{V.~Denisov}, \bibinfo{author}{V.~Zheltonozhskii},
  \bibinfo{author}{S.~Reshit'ko}, \bibinfo{journal}{Phys. At. Nucl.}
  \bibinfo{volume}{56} (\bibinfo{year}{1993}) \bibinfo{pages}{57}.
\bibitem[{{Chuvilskaya} et~al.(1999){Chuvilskaya}, {Seleznev}, {Shirokova}, and
  {Herman}}]{Chuvilskaya99-BRASP}
\bibinfo{author}{T.~V. {Chuvilskaya}}, \bibinfo{author}{Y.~G. {Seleznev}},
  \bibinfo{author}{A.~A. {Shirokova}}, \bibinfo{author}{M.~{Herman}},
  \bibinfo{journal}{Bull. Rus. Acad. Sci. Phys.} \bibinfo{volume}{63}
  (\bibinfo{year}{1999}) \bibinfo{pages}{825}.
\bibitem[{Ziegler et~al.(2010)Ziegler, Ziegler, and Biersack}]{SRIM}
\bibinfo{author}{J.~F. Ziegler}, \bibinfo{author}{M.~Ziegler},
  \bibinfo{author}{J.~Biersack}, \bibinfo{journal}{Nucl. Instrum. Methods Phys.
  Res., Sect. B} \bibinfo{volume}{268} (\bibinfo{year}{2010})
  \bibinfo{pages}{1818--1823}. \bibinfo{note}{http://www.srim.org}.
\bibitem[{Sz\"ucs et~al.(2014)Sz\"ucs, Kiss, and F\"ul\"op}]{Szucs14-AIPConf}
\bibinfo{author}{T.~Sz\"ucs}, \bibinfo{author}{G.~G. Kiss},
  \bibinfo{author}{Z.~F\"ul\"op}, \bibinfo{journal}{AIP Conf. Proc.}
  \bibinfo{volume}{1595} (\bibinfo{year}{2014}) \bibinfo{pages}{173}.
\bibitem[{McFarland(1991)}]{McFarland91-RR}
\bibinfo{author}{R.~C. McFarland}, \bibinfo{journal}{Radioactivity \&
  Radiochemistry} \bibinfo{volume}{2} (\bibinfo{year}{1991})
  \bibinfo{pages}{35}.
\bibitem[{Hubbell and Seltzer(2004)}]{NIST}
\bibinfo{author}{J.~Hubbell}, \bibinfo{author}{S.~Seltzer},
  \bibinfo{title}{Tables of {X-Ray} {Mass} {Attenuation} {Coefficients} and
  {Mass} {Energy-Absorption} {Coefficients} (version 1.4)},
  \bibinfo{year}{2004}. \bibinfo{note}{Avaliable:
  http://physics.nist.gov/xaamdi, National Institute of Standards and
  Technology, Gaithersburg, MD.}
\bibitem[{Rauscher and Thielemann(2001)}]{Rauscher01-ADNDT}
\bibinfo{author}{T.~Rauscher}, \bibinfo{author}{F.-K. Thielemann},
  \bibinfo{journal}{Atomic Data and Nuclear Data Tables} \bibinfo{volume}{79}
  (\bibinfo{year}{2001}) \bibinfo{pages}{47--64}.
\bibitem[{Rauscher(2012)}]{smaragd}
\bibinfo{author}{T.~Rauscher}, \bibinfo{title}{{Code} {SMARAGD}},
  \bibinfo{year}{2012}. \bibinfo{note}{Version 0.9.2s}.

\end{thebibliography}
\end{document}